\documentclass[twocolumn,english,aps,prx,superscriptaddress]{revtex4-1}
\usepackage[utf8]{inputenc}
\usepackage{color}
\usepackage{babel}
\usepackage{bm}
\usepackage{graphicx}
\usepackage{physics}
\usepackage{amsmath}
\usepackage{amssymb}
\usepackage{empheq}
\usepackage{times}
\usepackage{soul}
\usepackage{xr-hyper}

\usepackage[colorlinks=true,citecolor=blue]{hyperref}

\begin{document}

\title{Implementing two-qubit gates at the quantum speed limit}

\author{Joel Howard}
\affiliation{Department of Physics, Colorado School of Mines, Golden, Colorado 80401, USA}
\affiliation{Rigetti Computing, 775 Heinz Ave, Berkeley CA 94701}

\author{Alexander Lidiak}
\affiliation{Department of Physics, Colorado School of Mines, Golden, Colorado 80401, USA}

\author{Casey Jameson}
\affiliation{Department of Physics, Colorado School of Mines, Golden, Colorado 80401, USA}

\author{Bora Basyildiz}
\affiliation{Department of Computer Science, Colorado School of Mines, Golden, Colorado 80401, USA}

\author{Kyle Clark}
\affiliation{Department of Physics, Colorado School of Mines, Golden, Colorado 80401, USA}

\author{Tongyu Zhao}
\affiliation{Department of Physics, University of Colorado, Boulder, Colorado 80309, USA}

\author{Mustafa Bal}
\affiliation{Department of Physics, University of Colorado, Boulder, Colorado 80309, USA}
\affiliation{Superconducting Quantum Materials and Systems Center, Fermi National Accelerator Laboratory, Batavia, Illinois 60510, USA}

\author{Junling Long}
\affiliation{Department of Physics, University of Colorado, Boulder, Colorado 80309, USA}

\author{David P. Pappas}
\affiliation{National Institute of Standards and Technology, Boulder, Colorado 80305, USA}
\affiliation{Rigetti Computing, 775 Heinz Ave, Berkeley CA 94701}

\author{Meenakshi Singh}
\email{msingh@mines.edu}
\affiliation{Department of Physics, Colorado School of Mines, Golden, Colorado 80401, USA}

\author{Zhexuan Gong}
\email{gong@mines.edu}
\affiliation{Department of Physics, Colorado School of Mines, Golden, Colorado 80401, USA}

\begin{abstract}
The speed of elementary quantum gates, particularly two-qubit gates, ultimately sets the limit on the speed at which quantum circuits can operate. In this work, we experimentally demonstrate commonly used two-qubit gates at nearly the fastest possible speed allowed by the physical interaction strength between two superconducting transmon qubits. We achieve this quantum speed limit by implementing experimental gates designed using a machine learning inspired optimal control method. Importantly, our method only requires the single-qubit drive strength to be moderately larger than the interaction strength to achieve an arbitrary two-qubit gate close to its analytical speed limit with high fidelity. Thus, the method is applicable to a variety of platforms including those with comparable single-qubit and two-qubit gate speeds, or those with always-on interactions. We expect our method to offer significant speedups for non-native two-qubit gates that are typically achieved with a long sequence of single-qubit  and native two-qubit gates.
\end{abstract}

\maketitle

\section{Introduction} \vspace{-5pt}

Increasing the speed of elementary quantum gates boosts the ``clock speed" of a quantum computer. For noisy, intermediate-scale quantum computers \cite{Preskill2018} with finite coherence times \cite{Kjaergaard2019, TantalumQubits}, speeding up single- and two-qubit quantum gates also increases the circuit depth needed for solving useful computational problems \cite{Devoret2013, Supremacy2019}. In most experimental platforms, single-qubit  gates are achieved via electro-magnetic fields that drive individual qubit transitions. The maximum speed of these gates is often limited by the strength of the driving fields \cite{Frey2016, SingleQubitSL}. A two-qubit entangling gate, necessary for universal quantum gates, can however only operate at a speed proportional to the interaction strength between the qubits \cite{TwoQubitSL, trappedion_fast_theory1, trappedion_fast_theory2}, which is typically weaker than available single-qubit drive strengths and cannot be easily increased. 

Assuming a limited interaction strength, one can analytically obtain the maximum speed for any particular two-qubit gate in the limit of arbitrarily fast single-qubit gates \cite{Vidal2002,Kraus2001}. In practice, all single-qubit gates have finite speeds, and in platforms such as superconducting qubits, single-qubit gates may not be much faster than two-qubit gates due to limited anharmonicity \cite{Kandala2021,Stehlik2021,Moskalenko2022}. The speed limits of two-qubit gates in such scenarios have not been studied. Therefore, we seek to both theoretically and experimentally investigate these practical speed limits, which are not only relevant to the optimal design of quantum gates and quantum circuits, but also directly related to the speed limits of entanglement generation, a fundamental topic of high interest in quantum information theory, condensed matter physics, and black-hole physics \cite{LRbound,Guo2020,Hierarchy2020,Eldredge2017,scrambling}.

In this paper, we report a new method for designing two-qubit gates that are speed optimized, and we implement the gates experimentally using superconducting transmon qubits. We find that the protocol for achieving the fastest two-qubit gates in Refs.\,\cite{Vidal2002,Kraus2001} can be far from optimal with a finite single-qubit gate time. Our method differs from previous protocols \cite{Vidal2002,Kraus2001,Long2021} in that we apply single-qubit drives simultaneously with the two-qubit interaction, a crucial strategy for speed optimization. We optimize the pulse shapes of the single-qubit drives using a method that combines the well-known GRAPE algorithm \cite{OptimalControl,grape} with state-of-art machine learning techniques. Importantly, our method works as long as there exists a physical interaction between two qubits and one can drive each qubit with controllable pulse shape and phase. This applies to almost any platform suitable for quantum computing. Furthermore, we experimentally demonstrate that our method can achieve maximally entangling two-qubit gates, such as the CNOT gate, close to their analytical speed limits found in Refs.\,\cite{Vidal2002,Kraus2001} with modest single-qubit drive strengths and up to $98.3\%$ average gate fidelity determined from quantum process tomography. This largely eliminates the impractical assumption of infinitely fast single-qubit gates. The same applies to the SWAP gate we implemented, which is crucial for remote quantum gates but notoriously hard to achieve with a short gate sequence \cite{IBMprize}. We have also implemented a non-Clifford gate, the $\sqrt{\text{SWAP}}$ gate, close to its quantum speed limit at $97.0\%$ average gate fidelity.

We emphasize that the above-mentioned gates we implemented are all non-native gates, meaning that they cannot be achieved by evolving the static interaction Hamiltonian without single-qubit gates or time-dependent drives. These non-native gates are crucial for efficient implementation of many useful quantum circuits \cite{Abrams2020,Kivlichan2018,Babbush2018}. With our method, one could even realize an arbitrary two-qubit gate close to its theoretical speed limit. The conventional method of using a universal gate set to achieve a general $SU(4)$ unitary would instead require a sequence of up to 3 two-qubit gates and 12 single-qubit gates \cite{Vatan2004}, which is not only much slower in practice, but also likely of lower fidelity due to error accumulation. 

Although certain two-qubit gates with $99.5\%$ fidelity or higher have already been achieved with superconducting qubits \cite{Kandala2021,Stehlik2021,Moskalenko2022}, the lower fidelity gates reported here are largely due to limitations of our experimental hardware, as the theoretical gate fidelities associated with our demonstrated gates are all above $99.9\%$. The main purpose of this work is to demonstrate that the theoretical two-qubit gate speed limits can be largely achieved using a general optimal control method with minimal hardware requirement. Our speed-optimized gates achieve the same level of fidelity as that achieved previously on the same hardware with only fidelity optimization \cite{Long2021}, showing that our gate speed optimization does not rely on sacrificing gate fidelity. Moreover, our method is scalable to a large number of qubits provided that the two-qubit interactions can be switched on and off (such as via tunable couplers \cite{Liu2006}), since we can perform the speed optimization for each two-qubit gate independently. With most of the research on quantum gates focusing on improving gate fidelities, our work represents an orthogonal direction that aims to improve the fidelity of the whole quantum circuit \cite{Jurcevic2021} by optimizing the speed of any two-qubit gates.

\vspace{-5pt} \section{Experimental setup} \vspace{-5pt}

\begin{figure}[ht]
    \centering
    \includegraphics[width=\columnwidth]{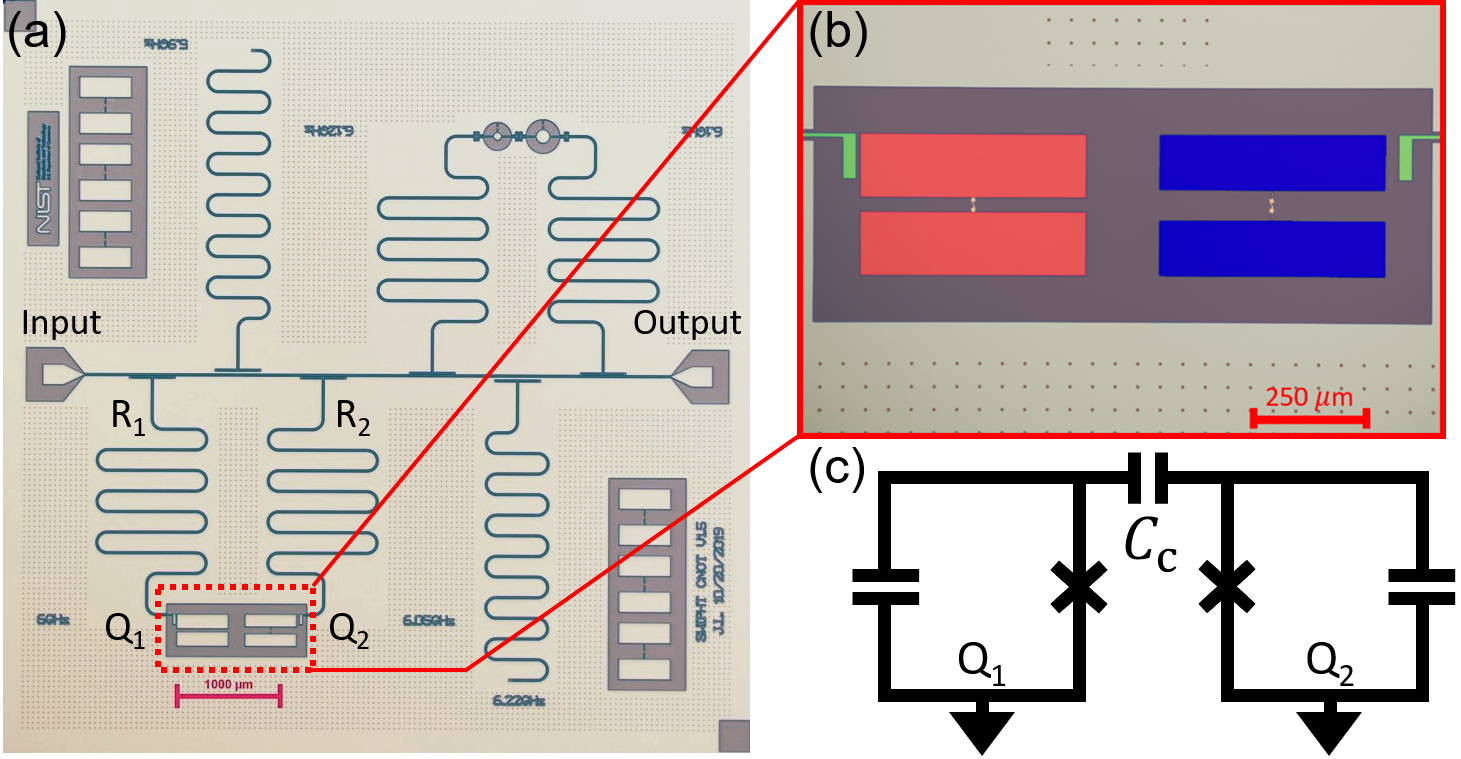}
    \caption{(a) Optical micrograph of the experimental chip including qubits, readout resonators, test Josephson junctions, and test resonators. (b) Zoomed-in view of the two floating qubits. Each qubit consists of two identical pads (red for the left qubit and blue for the right qubit) and a Josephson junction connecting the two pads. Each qubit is coupled to its own readout resonator (blue). (c) Grounded circuit model of the capacitively coupled qubits. }
\label{fig:TwoQubitChip}
\end{figure}

Our experimental platform consists of strongly-coupled fixed-frequency superconducting transmon qubits with static capacitive couplings in a hanger readout geometry \cite{SiddiqiHanger}. An intrinsic silicon substrate is used on which aluminum oxide tunnel junctions are fabricated via an overlap technique \cite{OverlapJunction}. The remaining circuit components are made of niobium. The full chip design and corresponding circuit model is shown in Fig.\,\ref{fig:TwoQubitChip}. The two transmon qubits' transition frequencies are $5.10\:\text{GHz}, 5.26\: \text{GHz}$, anharmonicities are $-270 \:\text{MHz},-320$ MHz, $T_1$ decay times are $40 \:\mu\text{s}, 21$ $\mu$s and $T^*_2$ decay times are $12\:\mu\text{s},10$ $\mu$s, respectively. In the rotating frame of the two qubit frequencies and assuming $\hbar=1$ from now on, the static Hamiltonian of the two qubits can be written as \cite{Long:2020,Long2021}
\begin{equation}
H_0 = g(\sigma_1^z+\sigma_2^z+\sigma_1^z \sigma_2^z)
\label{H0}
\end{equation}
where $g \approx 2\pi \times 1.75$ MHz represents a fixed Ising coupling strength between the qubits. To interact with the qubits, we deliver two microwave drives -- resonant with each qubit's transition frequency -- simultaneously through the feedline. Each of the drive fields contain two adjustable quadratures (X or Y), and can be described by the drive Hamiltonian.
\begin{equation}
    H_1(t)= \sum_{\gamma=x,y}\sum_{i=1,2}\Omega_i^{\gamma}(t)\tilde{\sigma}_i^{\gamma}
\label{H1}
\end{equation}
where $\Omega_i^{x,y}(t)$ denotes the Rabi frequency of the drive resonant with qubit $i$'s transition in the X or Y quadrature at time $t$.
For perfect single-qubit drives $\tilde{\sigma}_i^{\gamma}=\sigma_i^{\gamma}$. However, due to the strong Ising coupling between the two qubits in $H_0$, the drive strength on one qubit is dependent on the other qubit's state, resulting in $\tilde{\sigma}_1^{\gamma}=\sigma^{\gamma}\otimes (\dyad{0}{0}+r_2\dyad{1}{1})$ and $\tilde{\sigma}_2^{\gamma}=(\dyad{0}{0}+r_1\dyad{1}{1}) \otimes \sigma^{\gamma}$, with $r_1 \approx 1.1$ and $r_2 \approx 0.7$ for our current chip. We note that with a weaker coupling strength or with a tunable coupler \cite{Houck2019, MITCoupler}, both $r_1$ and $r_2$ can be made closer to or equal to 1.

\section{Analytical speed limit}

In the limit of arbitrarily strong single-qubit drives, i.e. $\Omega_{\text{max}}\equiv\max{|\Omega_{1,2}^{x,y}(t)|}\rightarrow \infty$, one can derive an analytical speed limit for any target two-qubit unitary with the above-mentioned static Hamiltonian $H_0$ and control Hamiltonian $H_1$. Note that in this limit, the speed limit is only well defined when $r_1=r_2=1$, since otherwise $H_1$ will lead to arbitrarily strong interactions. As detailed in \cite{Kraus2001}, any two-qubit target unitary $U$ can always be decomposed as
\begin{equation}
    U=(U_1 \otimes U_2) U_d (V_1 \otimes V_2), \quad U_d = e^{-i\sum_{\gamma} \lambda_{\gamma} \sigma^{\gamma} \otimes \sigma^{\gamma}}
\label{Udecomp}
\end{equation}
where $U_1, V_1$ ($U_2, V_2$) are some single-qubit gates on the first (second) qubit, $\gamma=x,y,z$, and $\lambda_{\gamma}\in [-\frac{\pi}{4},\frac{\pi}{4}]$. $\{\lambda_x,\lambda_y,\lambda_z\}$ form a canonical vector that uniquely specifies any given two-qubit gate $U$ up to single-qubit rotations, and their exact values can be obtained with the knowledge of $U$ based on Ref.\,\cite{Vidal2002}. \color{black}
To obtain the analytical speed limit, we assume that single qubit gates are arbitrarily fast and thus take a negligible amount of time. The total gate time for implementing $U$ therefore reduces to the time spent on realizing $U_d$, which is responsible for any entanglement generation. Based on $H_0$, together with instantaneous single-qubit rotations, $U_d$ can be realized with a minimum time of:
\begin{equation}
    T_{\text{min}}=\frac{|\lambda_x| + |\lambda _y| +|\lambda_z|}{g}= \begin{cases}
        \pi/(4g) & \text{CNOT}\\
        3\pi/(4g) & \text{SWAP}\\
        3\pi/(8g) & \sqrt{\text{SWAP}}
    \end{cases}
\label{Tmin}
\end{equation}
Eq.\,\eqref{Tmin} is the analytical speed limit for a two-qubit gate $U$. Its values for the CNOT, SWAP, and $\sqrt{\text{SWAP}}$ gate are shown above and derived in Appendix B.

\vspace{-5pt} \section{Optimal control method} \vspace{-5pt}

\begin{figure}
    \includegraphics[width=\columnwidth]{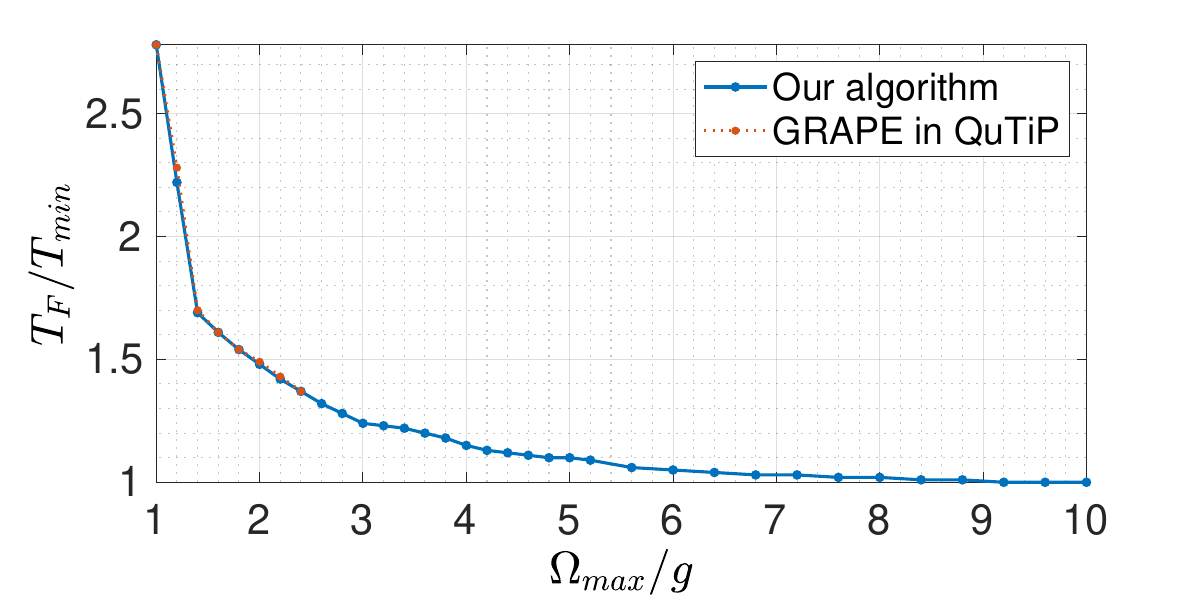}
    \caption{The minimum time $T_F$ (in units of $T_{\text{min}}$) it takes to achieve a CNOT gate of $F>99$\% as a function of $\Omega_{\text{max}}$ (in units of $g$) using either our optimization algorithm (blue) or the GRAPE algorithm in QuTiP (red). Both algorithms use $16$ segments of the drive pulses and $200$ random restarts. The GRAPE algorithm in QuTiP fails to reach $F>99\%$ for larger $\Omega_{\text{max}}$ values, where the gate time approaches its theoretical limit $T_{\text{min}}$.}
    \label{fig:OptCNOT}
\end{figure}

In practice, single-qubit gate speeds are limited by finite drive strengths and the analytical speed limit in Eq.\,\eqref{Tmin} does not apply. To our best knowledge, no analytical speed limit has been found with finite single-qubit gate time. In this case, we can no longer rely on the decomposition in Eq.\,\eqref{Udecomp} to reduce the problem to just finding the minimum time in implementing $U_d$. The true time-optimal protocol may not feature the structure of such decomposition and is challenging to find analytically for a general target gate. In fact, if we still follow Eq.\,\eqref{Udecomp} for realizing a target two-qubit gate, the resulting gate time can be much longer than $T_{\text{min}}$. To realize universal single-qubit gates needed in Eq.\,\eqref{Udecomp}, we use a two-axis gate (TAG) protocol first developed in Ref.\,\cite{Long2021}, which employs an analytically obtained 3-segment drive pulse for $\Omega_{1,2}^{x,y}(t)$ to exactly cancel the effects of static interaction for any values of $r_1$ and $r_2$. A single-qubit gate implemented this way has a gate time of at least $\pi/(2g)$ \cite{Long:2020}. Apart from the native CZ gate that can be directly realized via an evolution of $H_0$ over a time $t=\pi/(4g)\approx 71.4$ns \cite{Collodo2020,Long2021, Barron2020}, any two-qubit gate design that involves the use of single-qubit gate(s) realized via TAG requires a gate time of at least $T_{\text{min}}+\pi/(2g)$ (since single-qubit gates cannot shorten $T_{\text{min}}$), and is thus far from optimal.

Consequently, to approach the analytical speed limit with finite $\Omega_{\text{max}}$ (or finite single qubit gate time), we adopt an alternative approach that avoids the use of any single-qubit gate and generate the target two-qubit gate directly. Specifically, we directly optimize the pulse shapes $\Omega_{1,2}^{x,y}(t)$ in our control Hamiltonian $H_1(t)$ in order to minimize the gate time for achieving a certain target gate with sufficiently high fidelity. For a given set of pulse shape functions $\Omega_{1,2}^{x,y}(t)$, we numerically find the evolution operator 
\begin{equation}\label{U}
\mathcal{U}=\mathcal{T} e^{-i \int_0^{T} [H_0 +H_1(t)]dt}
\end{equation}
where $\mathcal{T}$ denotes the time ordered integral and $T$ denotes the total evolution time. Note that $\mathcal{U}$ can achieve any two-qubit gate with properly engineered pulse shapes, as the Hamiltonian and the commutators of the Hamiltonian at different times span all $SU(4)$ generators. 

Next, we calculate the average gate fidelity  between the target unitary $U$ and the evolution operator $\mathcal{U}$ using \cite{NielsenFidelity}
\begin{equation}
    F= \frac{1}{5} + \frac{1}{20}\sum_j \Tr(U U_j U^{\dagger} \mathcal{U} U_j \mathcal{U}^{\dagger})
\label{AF}
\end{equation}
where $U_j \in \{\sigma^{\gamma}\otimes \sigma^{\gamma^{\prime}}\}$ and $\sigma^{\gamma} \in \{\sigma^x,\sigma^y,\sigma^z,I\}$. For efficient numerical optimization, we will assume $\Omega_i^{\gamma}(t)$ is an $M$-segment piece-wise function, i.e. $\Omega_i^{\gamma}(t)=\Omega_{i,m}^{\gamma}$ for $t\in [\frac{m-1}{M}T,\frac{m}{M}T]$, where $m$ indexes the segments sequentially. Our goal is to maximize $F$ over all possible values of $\{\Omega_{i,m}^{\gamma}\}$ subjected to the constraints $|\Omega_{i,m}^{\gamma}|\le \Omega_{\text{max}}$ for a given time $T$. The numerical speed limit $T_F$ is then defined as the minimum $T$ that can achieve $F>1-\epsilon$, where $\epsilon$ is the infidelity we can tolerate (set to $1\%$ in the following).

Since $F$ is a highly nonlinear function of \{$\Omega_{i,m}^{\gamma}$\}, simple numerical optimization methods will not work well in finding the global maximum of $F$. Here we develop a new method that combines the standard GRAPE algorithm \cite{grape} with state-of-art machine learning techniques. Using the backward propagation method in the widely used machine learning library PyTorch \cite{pytorch}, we  calculate the gradients of $F$ over each pulse parameter $\Omega_{i,m}^{\gamma}$ automatically. We then perform a stochastic gradient descent (SGD) algorithm with the Nesterov Momentum method \cite{nesterov1983method} to maximize $F$ over the pulse parameters. To avoid obtaining only a local maximum for $F$, we repeat each gradient descent process with $200$ random seeds used for both initialization and SGD, and then select the global maximum among all repetitions. Further increasing the number of random seeds does not lead to noticeable improvement in maximizing $F$, showing that the optimization has converged.

To benchmark our numerical optimization method, we choose the target gate to be the CNOT gate and find the above-mentioned numerical speed limit $T_F$ for $F=99\%$ as a function of $\Omega_{\text{max}}$. We set $M=16$, which allows the calculation to be done within a few hours on a small HPC cluster, and larger $M$ does not lead to noticeable improvements. As shown in Fig.\,\ref{fig:OptCNOT}, we clearly see that as $\Omega_{\text{max}}/g$ increases, $T_F$ approaches the analytical speed limit $T_{\text{min}}$, indicating that the optimization succeeded in reaching the theoretical speed limit. Importantly, the maximum single-qubit drive strength $\Omega_{\text{max}}$ does not need to be significantly larger than the interaction strength $g$ to get close to the analytical speed limit. For example, setting $\Omega_{\text{max}}=3g$ already gives us a minimum gate time of $1.24T_{\text{min}}$ with $F>99\%$. We also compare our method with the standard GRAPE algorithm in the widely used QuTiP software \cite{qutip}. With the same number of iterations and random initializations, the GRAPE algorithm in QuTiP can closely match our optimization results for $\Omega_{\text{max}}/g<2.5$. However, it fails to achieve $F>99\%$ for $\Omega_{\text{max}}>2.5g$, and the same happens even with double the number of iterations or random initializations. This is likely because the algorithm struggles at escaping local minima due to a larger parameter space for a larger value of $\Omega_{\text{max}}$.

In Appendix C, we show that our optimization method also works well for different interaction Hamiltonians, such as the flip-flop interaction common in superconducting qubit systems. With such interaction, the speed optimization can significantly speed up CNOT and CZ gates since they can operate as fast as the native iSWAP gate. In contrast, most existing experiments with flip-flop interacting Hamiltonians report much slower CZ gates compared to iSWAP gates \cite{Kandala2021,Moskalenko2022,Sung2021}.

\vspace{-5pt} \section{Experimental results} \vspace{-5pt}

\begin{figure*}
    \centering
    \includegraphics[width=0.33\textwidth]{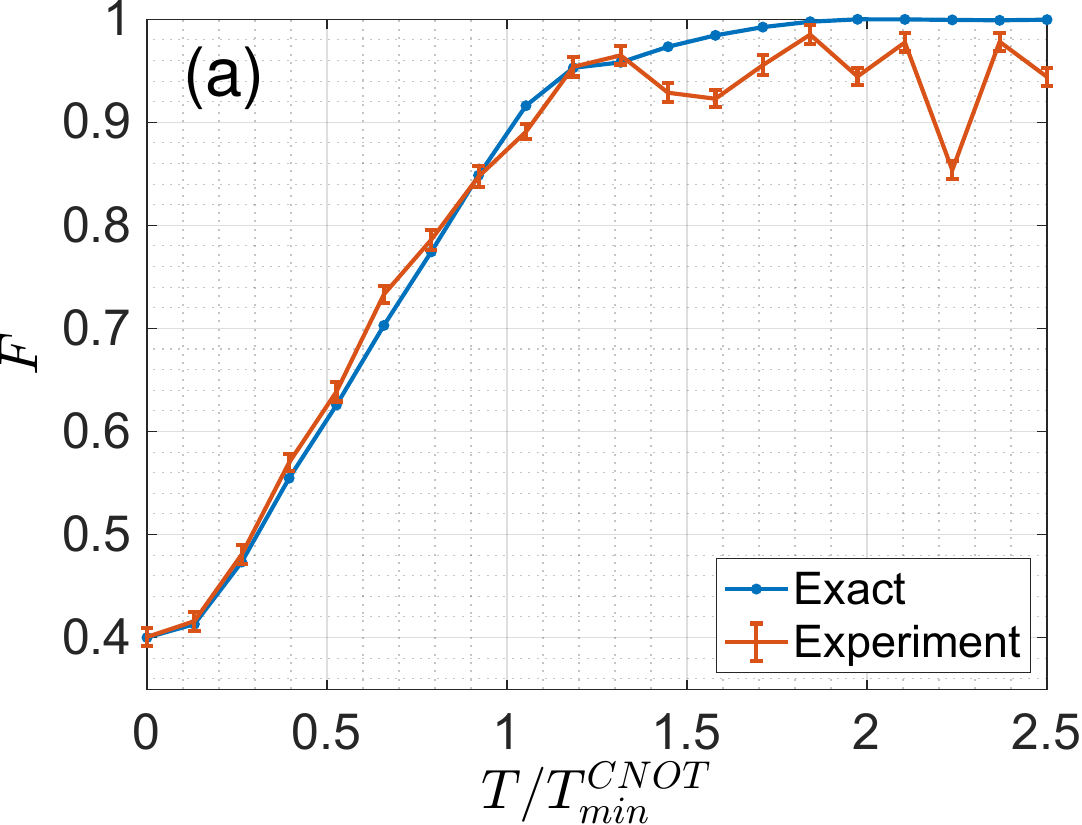}
    \includegraphics[width=0.33\textwidth]{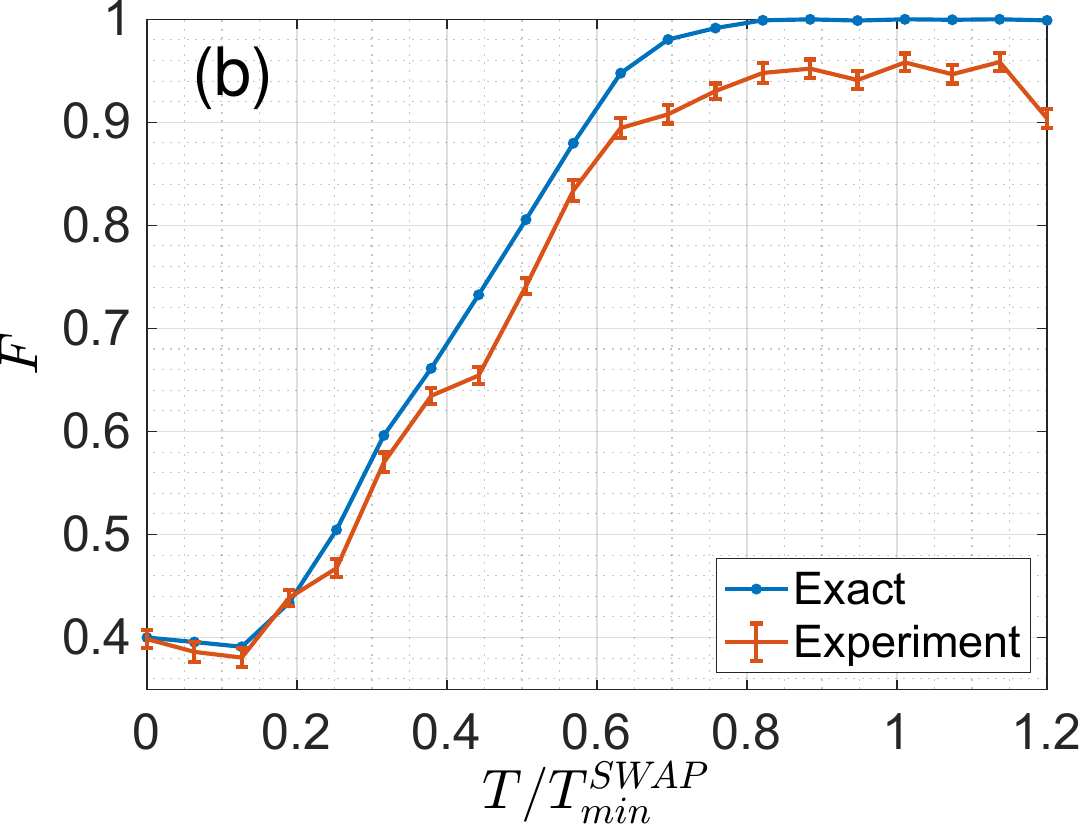}  
    \includegraphics[width=0.33\textwidth]{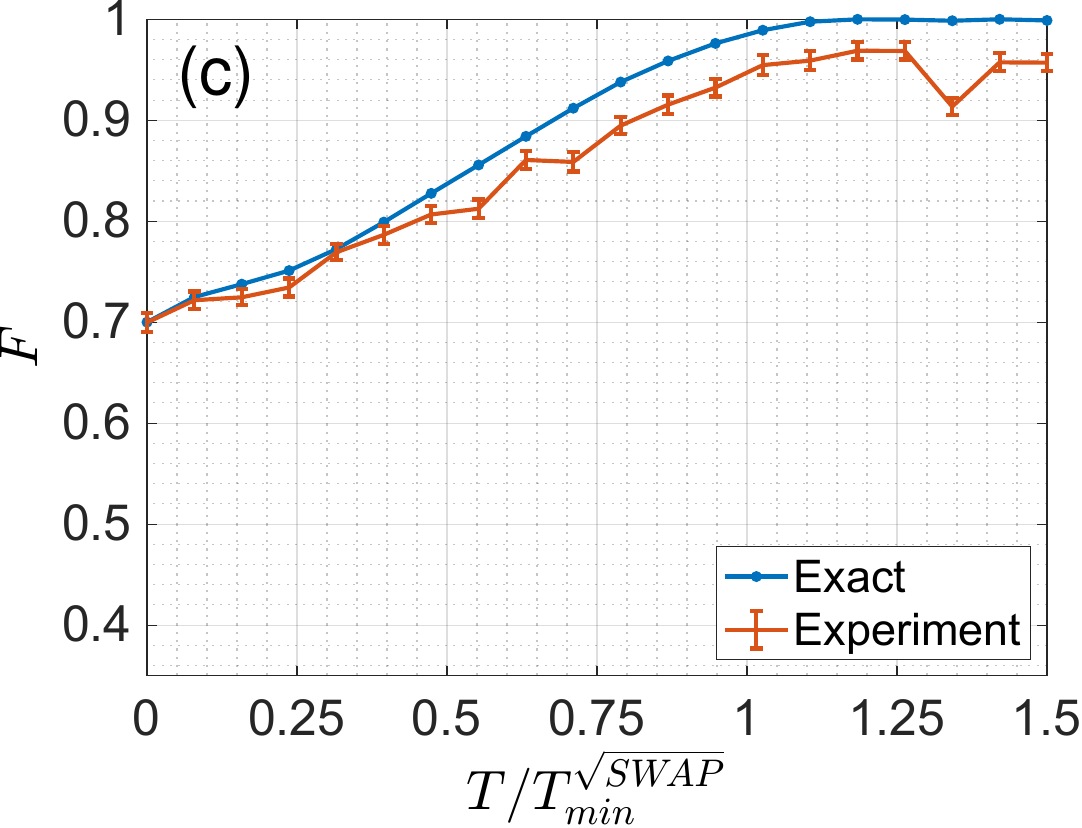}
  
    \caption{Experimental measurements of the average gate fidelity $F$ using optimized 4-segment drive pulses , with the target gate being (a) CNOT, (b) SWAP, and (c) $\sqrt{\text{SWAP}}$. The red curves represent experimental measurements while the blue curves represent the exact numerical calculation of $F$ without considering any experimental error. $\Omega_{\text{max}}=6$MHz for the CNOT gate and $\Omega_{\text{max}}=5$MHz for SWAP and $\sqrt{\text{SWAP}}$ gate. The error bars represent a upper bound on the statistical error of the mean for 500 repeated measurements at each point.}
    \label{fig:Experiment}
\end{figure*}

We now proceed to demonstrate the speed limits of the two-qubit CNOT, SWAP, and $\sqrt{\text{SWAP}}$ gates experimentally. The procedure for this is as follows. First, for each gate, the total evolution time $T$ is varied from $0$ to $\gtrsim T_{\text{min}}$ in 20 steps and the optimized pulse sequence is obtained numerically for each value of $T$.  Next, this pulse sequence is applied to the transmon qubits experimentally by modulating the microwave drive signals. Finally, the average gate fidelity $F$ is measured at time $T$ by performing a quantum process tomography (QPT) \cite{GroveTomography}. Our QPT involves applying 36 different pre-rotations to an initial state with both qubits in the state $\ket{0}$, applying the optimized pulse sequence for time $T$, and then measuring 9 different Pauli operator (see Appendix D for details), resulting in 324 different experimental protocols, each of which is further repeated 500 times to ensure low statistical errors. After correcting the state preparation and measurement (SPAM) errors as well as performing a maximum likelihood estimation to ensure a completely positive and trace-preserving quantum map (see Appendix D for details), the QPT allows us to find a Pauli transfer matrix \cite{Chow2012} for the corresponding quantum process, which can be further used to infer $F$ (Appendix D). This process allows us to find the value of $T$ above which we can get sufficiently high gate fidelity. Such $T$ is the experimental speed limit for the target gate.

There are several experimental limitations in this procedure. First, as strong microwave drives can heat up the superconducting qubits and cause decoherence, we only send microwave pulses of at most $2\pi \times 6$ MHz in Rabi frequency, roughly 3 times the coupling strength $g$. But as we have shown in Fig.\,\ref{fig:OptCNOT}, this limitation should not prevent us from getting close to the analytical speed limit. A more noticeable limitation is that we can only generate smoothly varying pulse shapes that approximate the segmented (and thus discontinuous) pulse shapes used in the numerical optimization. As the number of segments $M$ increases, this approximation deteriorates while the gate speed increases (and eventually converges). For our setup, we choose $M=4$ for the experiment as a sweet spot for balancing the error and speed. We note that this limitation can be addressed by numerically optimizing smooth pulse shape functions (such as a train of Gaussian envelopes), although such optimization is more resource intensive. Finally, with $r_1,r_2\ne 1$ experimentally, our single-qubit drives will induce a small amount of extra interaction that would in principle allow us to go above the analytical speed limit for sufficiently large $\Omega_{\text{max}}$. Our numerical optimizer accounts for this artifact. The amount of speedup over the scenario of $r_{1,2}=1$ varies for different target gates.

Our experimental results are shown in Fig.\,\ref{fig:Experiment}. The measured gate fidelity $F$ (red curves) closely matches the one obtained from the numerical simulation of the experiment with no error (blue curves). The deviations between the two grow as the gate fidelity gets close to $1$ for reasons we discuss in the next section. For the CNOT gate, we were able to achieve $F\approx96.5\%$ experimentally with a gate time of $T=93.7\text{ns}\approx1.32 T_{\text{min}}$ (Fig.\,\ref{fig:Experiment}a). We emphasize that this outperforms the CNOT gate implemented using the SWIPHT protocol \cite{swipht_theory} performed on the same hardware ($F\approx94.6\%$ for a gate time of $1.87T_{\text{min}}$ \cite{Long2021}), which is protocol designed specially for our hardware. The highest fidelity we achieve is $F\approx 98.3\%$ at time $T\approx1.84T_{\text{min}}$.

For the SWAP and $\sqrt{\text{SWAP}}$ gates, the extra interactions caused by non-unity $r_1$ and $r_2$ values have a more noticeable effect in speeding up the gates. For the SWAP gate (Fig.\,\ref{fig:Experiment}b), we obtain an experimental gate fidelity of $F\approx95.9\%$ at $T=216\text{ns}\approx 1.01T_{\text{min}}$, where theoretically $F\approx 99.997\%$. A SWAP gate with such a short time is hard to achieve via a gate sequence using a typical universal gate set, making our method particularly useful given the importance of SWAP gates in many quantum algorithms \cite{IBMprize}. For the $\sqrt{\text{SWAP}}$ gate (Fig.\,\ref{fig:Experiment}c), we obtain an experimental gate fidelity of $F\approx97.0\%$ at $T=126\text{ns}\approx 1.18T_{\text{min}}$, with $F\approx 99.999\%$ in theory. 

For all gates, the demonstrated experimental speed limits are reasonably close to the analytical speed limits. We note that the fidelities achieved here are lower than state-of-the-art due to limitations of the hardware platform (discussed in the next section) and not due to the optimal control algorithm. Even without optimal control, the fidelities obtained on this setup are close to or lower than what we are getting here \cite{Long2021}.

\vspace{-5pt} \section{Error analysis} \vspace{-5pt}

\begin{figure} 
    \includegraphics[width=0.9\columnwidth]{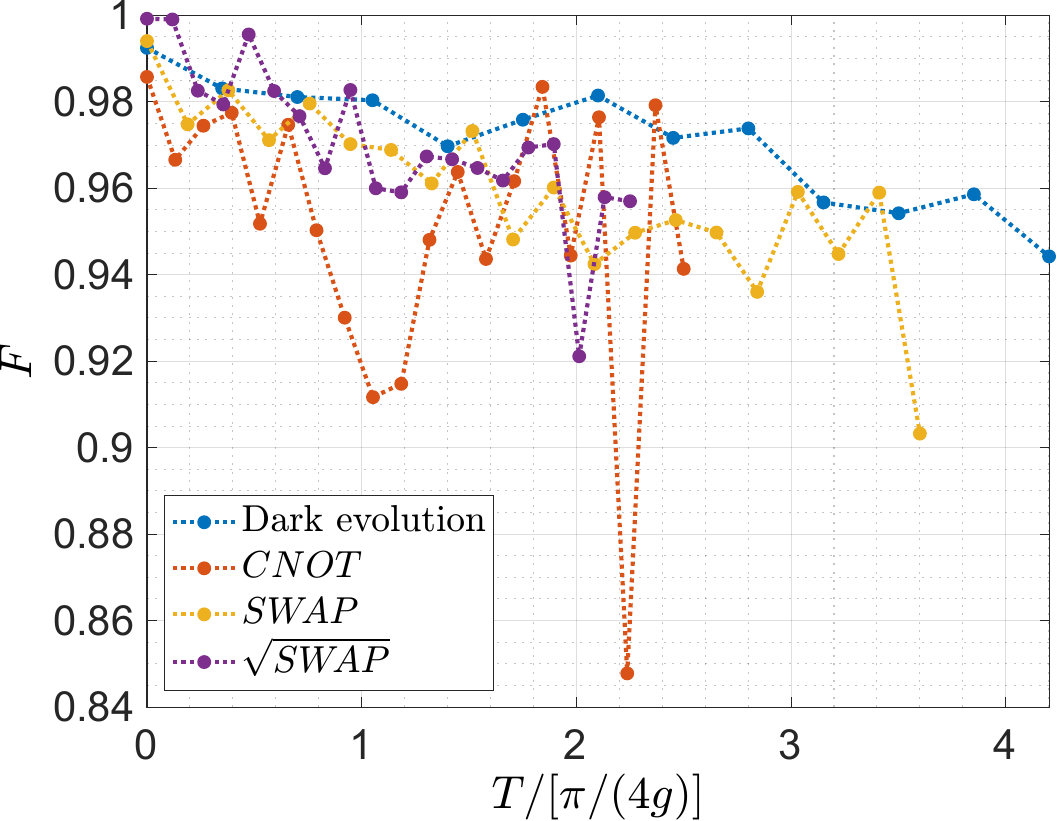}

    \caption{Fidelity $F$ between the experimental quantum process (characterized by the QPT) and the corresponding exact time evolution operator in Eq.\,\eqref{U} using the optimized pulse shapes for a given gate time $T$ with the target gate being CNOT, SWAP, or $\sqrt{\text{SWAP}}$. The blue curve represents the fidelity between the experimental evolution without the drives (i.e. dark evolution) and the ideal evolution operator of $e^{-iH_0 T}$.} \label{fig:error}
\end{figure}

We have calculated the fidelity between the experimental process and the exact time evolution operator $\mathcal{U}$ in Eq.\,\eqref{U} for each point in Fig.\,\ref{fig:Experiment}, which is in general $>95\%$ (see Fig.\,\ref{fig:error}) As seen from Fig.\,\ref{fig:Experiment}, the experimental errors get larger at large values of $T$. This is possibly due to the following reasons. 

First, the qubits decohere as time increases. This is evidenced by our measurement of a dark evolution (i.e. with the drive Hamiltonian $H_1$ turned off) process fidelity that drops from $\approx 99.3\%$ to $\approx 96.3\%$ from $T=0$ to $T=3\pi/(4g)$ (the theoretical minimum time for the SWAP gate), as shown in Fig.\,\ref{fig:error}. This large loss in fidelity is unrelated to optimal control or errors in the control pulses, and likely results from finite $T_1$ time of the qubits, measurement errors, and low-fidelity (about $98\%$ on average) single-qubit gates used in our QPT \cite{Long:2020,Long2021}. With better hardware designs, these errors can be largely eliminated. For example, single-qubit gates with $>99.9\%$ fidelities have  already been achieved with superconducting qubits \cite{Manenti2021}.

Second, when $T$ is large enough to allow the numerically optimized $F$ to approach 1, imperfect calibration or fluctuations on the microwave drive amplitudes or phases tend to create a larger discrepancy between the experiment and the theory, as we have discussed in detail in Appendix E. This can account for up to $0.1 \%$ loss in fidelity for $\approx 1 \%$ deviations in pulse shapes.

\begin{figure}
    \centering
    \includegraphics[width=0.9\columnwidth]{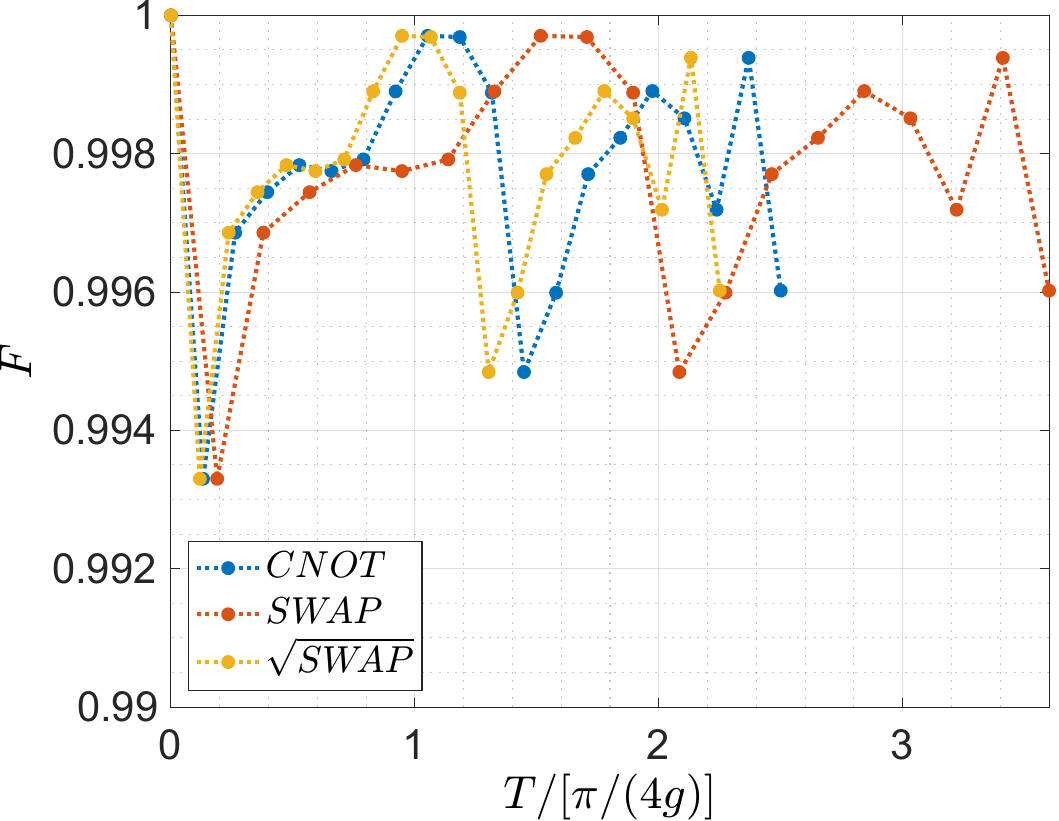}
    \caption{Average fidelity $F$ between the evolution operators calculated using the Hamiltonian in Eq.\,\eqref{Hfull} and using Eq.\,\eqref{U} for the speed-optimized CNOT, SWAP, and $\sqrt{\text{SWAP}}$ gates shown in Fig.\,\ref{fig:Experiment}.} \label{fig:syserrors}
\end{figure}

Finally, there are also systematic errors coming from the leakage to higher excited states (in particular the $|2\rangle$ state for each transmon), cross talk between the drives of each qubit, rotating wave approximations, and the deviation of the experimental pulse shapes from the ideal square waves used in our numerical optimization. We can characterize these errors with the following Hamiltonian in the lab frame for two qutrits:
\begin{align}\label{Hfull}
     &H(t) = \sum_m E_m \dyad{m}{m} \nonumber\\ &+ \frac{1}{2}\sum_{i=1,2}\sum_{m\ne m^{\prime}}\left(d_{m, m^{\prime}}\mathcal{E}_i(t)\dyad{m}{m^{\prime}}e^{-i\omega_i t}+\text{h.c.}\right)
\end{align}
where $m\in\{00,01,10,11,02,20,12,21,22\}$ labels the energy eigenstates of the static Hamiltonian, and $d_{m,m^{\prime}}$ represents the dipole moment for the transition between states $|m\rangle$ and $|m^{\prime}\rangle$. $\mathcal{E}_1(t)$ and $\mathcal{E}_2(t)$ denote the electric fields of the two microwave drives we applied at frequencies $\omega_1 = E_{10} -E_{00}$ and $\omega_2 = E_{01} - E_{00}$ respectively. Their values are set by the actual experimentally applied electric fields that follow the pulse shapes from our optimization method but have finite rising/lowering edges between different pulse segments. The Hamiltonian in Eq.\,\eqref{Hfull} then fully models the leakage outside the qubit subspace, the cross talk between the two drive fields, and realistic pulse shapes without rotating wave approximations. We then calculate the fidelity between the exact evolution operator of this Hamiltonian and the one in Eq.\,\eqref{U} (which was used for Figs.\,\ref{fig:Experiment}-\ref{fig:error}). As shown in Fig.\,\ref{fig:syserrors}, these errors only add up to about $0.3\%$ infidelity on average. And since these errors are explicitly modelled by the Hamiltonian in Eq.\,\eqref{Hfull}, we can further minimize their impact to gate fidelities using the same optimization method we built. However, this is beyond the scope of this work as our experiment hardly benefits from such effort due to other error sources being dominating.

\vspace{-5pt} \section{Conclusion and Outlook} \vspace{-5pt}

There are primarily three advancements made in this work. First, we have studied the speed limits for two-qubit gates under realistic experimental conditions and shown that these limits are close to the analytical speed limits derived under ideal conditions. Second, we have developed an optimal control algorithm to generate a realistic pulse sequence to achieve these speed limits. Our algorithm performs better than a standard GRAPE algorithm and can be used to design speed-optimized two-qubit gates in a variety of quantum computing platforms with different types of interactions. It can offer significant speedups for non-native two-qubit gates especially when single-qubit gate times are not negligible. Finally, we have experimentally demonstrated the quantum speed limits for various two-qubit gates using superconducting qubits.

We have also carefully characterized the error sources for our experimental gates. Most of the gate errors come from characterization/calibration errors, imperfect measurements, qubit decoherence, and low-fidelity single-qubit gates. While the strong drive pulses in the optimal control could lead to more leakage and cross-talk errors, we show that these errors only add up to about $0.3\%$ infidelity on average, and they can be further mitigated by optimizing a more accurate Hamiltonian. It is also worth pointing out that by optimizing the speed of two-qubit gates, errors from qubit decoherence will be suppressed. We therefore expect our method to be able to improve the fidelity of the whole quantum circuit.

An important future direction is to generalize this work to a multi-qubit scenario where additional qubits are used to speed up a two-qubit gate. Previous work has shown that significant scaling speedups may be obtained in performing remote quantum gates or preparing useful many-body entangled states \cite{Eldredge2017,Hierarchy2020} with long-range interacting qubits. However, questions regarding the speed limit of entangling gates when interactions are strongly long-ranged are still largely open \cite{Guo2020}. Such interactions play important roles in quantum information scrambling \cite{scrambling} and the development of fully-connected quantum computers \cite{MonroePNAS}. Another interesting direction is to study the speed limit of entangling gates when higher excited states outside the qubit subspace are utilized \cite{anharmonicSpeedLimit}, where experimental and analytical results are both lacking.

\vspace{-5pt}
\begin{acknowledgments}
\vspace{-5pt}
This work is jointly supervised by MS and ZXG. We thank NIST Boulder for hosting the experiment and the HPC center at Colorado School of Mines for providing computational resources. We acknowledge funding support from the NSF RAISE-TAQS program CCF-1839232, the NSF Triplets program DMR-1747426, the NSF NRT program DGE-2125899, and the W. M. Keck Foundation.
\end{acknowledgments}

\appendix

\vspace{-5pt} 
\section{Experimental Hardware, Calibration, and Characterization}
\vspace{-5pt} 

\begin{figure*}
    \centering
    \includegraphics[width=0.8\textwidth]{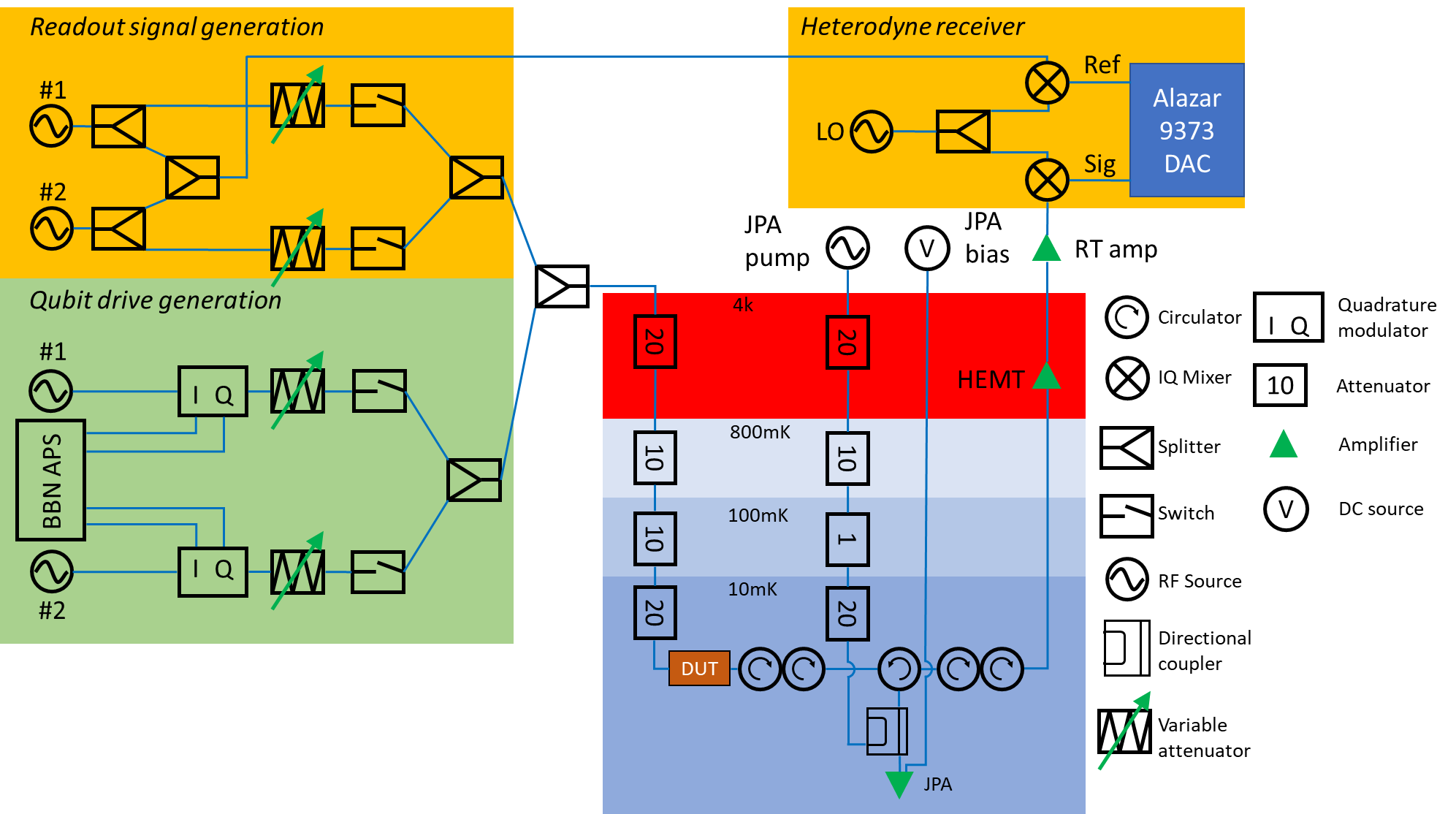}
    \caption{Our experimental setup composed of qubit drives and heterodyne state readout. Each qubit drive (green section) is shaped via quadrature modulation by a BBN APS1 and Polyphase Microwave AM4080A. Readout (orange) consists of digitally locking in the signal passing through the device with a reference and extracting I/Q shifts used to classify the ground/excited states.}
    \label{ExpSetup}
\end{figure*}
Our experimental device is operated at 10mK in an Bluefors LD dilution refrigerator. Full schematics of the experimental setup are shown in Fig.\,\ref{ExpSetup}. All qubit drive and readout microwave tones are delivered via the feedline, which has an output amplification chain of a Raytheon BBN Josephson parametric amplifier (JPA) preamp at base, high-electron-mobility transistor (HEMT) amplifier at 4K, and a high-gain room-temperature amplifier.

Each experimental cycle consists of a state initialization, a time evolution under the engineered Hamiltonian flanked by process tomography rotations \cite{GroveTomography} and followed by a heterodyne state readout (see Fig.\,\ref{ExpSetup}). The state initialization occurs by waiting $500\mu\text{s} \approx 12T_1$ between two experimental cycles, which is long enough to guarantee that each qubit is in the $|0\rangle$ state. All gates consist of microwave tones from a Holzworth HS9008B pulse shaped by a BBN arbitrary pulse sequencer (APS) quadrature modulation scheme. Readout consists of a simultaneous 2$\mu$s probe (Agilent N5183Ms) of the two readout resonators to detect shifts in their frequencies due to their respective qubit states. The I/Q components of the readout signal shift are extracted via down conversion and a digital lock-in routine with a reference tone. They are then used to identify the two-qubit states as  $\ket{00},\ket{01},\ket{10},\ket{11}$ via a classification algorithm using support vector machines. Total state preparation and measurement errors, quantified by the basis-state preparation confusion matrix \cite{GroveTomography} stayed under $5\%$, with the errors dominated by readout errors associated with qubit state relaxation during the measurement.

The computational subspace spectrum of the transmons was determined via a combination of spectroscopy (directly probing excitations with a $10 \mu s$ square pulse) and Ramsey experiments (driving 2 MHz off-resonant, running a typical Ramsey sequence, and noting the deviation of the fitted frequency from 2MHz) \cite{Ramsey1950}. The strength of the drive fields on the qubits was inferred from the frequency of Rabi oscillations of the excited state population incurred by driving at uniform strength for a linearly increasing duration. The linearity of the pulse shaping quadrature channels on the pulse sequencer was characterized by measuring the Rabi oscillation frequency resulting from a sweep over pulse amplitudes, analyzing it via Fourier filtering \cite{Long:2020}, and correcting for it at the software level.

\vspace{-5pt} 
\section{Obtaining Analytical Speed Limits}
\vspace{-5pt} 

We provide details on how to obtain the analytical speed limit $T_{\text{min}}$ defined in Eq.\,(4) of the main text. Given a two-qubit unitary operator $U$, the key step is to find the decomposition of $U$ into $(U_1 \otimes U_2) U_d (V_1 \otimes V_2)$ where $U_d = e^{-i\sum_{\gamma=x,y,z} \lambda_{\gamma} \sigma^{\gamma} \otimes \sigma^{\gamma}}$ with $\lambda_{x,y,z}\in [-\frac{\pi}{4},\frac{\pi}{4}]$, and $U_1, V_1$ ($U_2, V_2$) are some single-qubit gates on the first (second) qubit. This decomposition is non-trivial, and the detailed procedure can be found in Ref.\,\cite{Kraus2001}. Here we provide the results of the decomposition for the three target gates we studied in Table\,\ref{tab:decomp}. The values of $\lambda_{x,y,z}$ directly lead to $T_{\text{min}}$ values for the three target gates shown in Eq.\,(4) of the main text. Note that an overall phase difference is tolerated for the decomposition of $U$.

\begin{table}[h]
\begin{tabular}{|c|c|c|c|}
\hline 
 & CNOT  & SWAP  & $\sqrt{\text{SWAP}}$ \tabularnewline
\hline 
\hline 
$U_{1}$  & $\frac{1}{\sqrt{2}}\begin{pmatrix}1 & -1\\
1 & 1
\end{pmatrix}$  & $\begin{pmatrix}1 & 0\\
0 & 1
\end{pmatrix}$  & $\begin{pmatrix}0 & 1\\
-1 & 0
\end{pmatrix}$ \tabularnewline
\hline 
$U_{2}$  & $\begin{pmatrix}1 & 0\\
0 & 1
\end{pmatrix}$  & $\begin{pmatrix}1 & 0\\
0 & 1
\end{pmatrix}$  & $\begin{pmatrix}1 & 0\\
0 & -1
\end{pmatrix}$ \tabularnewline
\hline 
$V_{1}$  & $\frac{1}{\sqrt{2}}\begin{pmatrix}1 & i\\
-1 & i
\end{pmatrix}$  & $\begin{pmatrix}1 & 0\\
0 & 1
\end{pmatrix}$  & $\begin{pmatrix}0 & 1\\
-1 & 0
\end{pmatrix}$ \tabularnewline
\hline 
$V_{2}$  & $\frac{1}{\sqrt{2}}\begin{pmatrix}1 & -1\\
-1 & 1
\end{pmatrix}$  & $\begin{pmatrix}1 & 0\\
0 & 1
\end{pmatrix}$  & $\begin{pmatrix}1 & 0\\
0 & -1
\end{pmatrix}$ \tabularnewline
\hline 
$\lambda_{x}$  & $\frac{\pi}{4}$  & $\frac{\pi}{4}$  & $\frac{\pi}{8}$ \tabularnewline
\hline 
$\lambda_{y}$  & 0  & $\frac{\pi}{4}$  & $-\frac{\pi}{8}$ \tabularnewline
\hline 
$\lambda_{z}$ & 0 & $\frac{\pi}{4}$ & $-\frac{\pi}{8}$ \tabularnewline
\hline 
\end{tabular}
\caption{Detailed decompositions of the CNOT, SWAP, and $\sqrt{\text{SWAP}}$ gates based on Eq.\,\eqref{Udecomp}.} \label{tab:decomp}

\end{table}

\vspace{-5pt} 
\section{Optimization for different static Hamiltonians}
\vspace{-5pt}

To demonstrate the universality of our optimal control method, here we apply it to two different static Hamiltonians commonly seen in superconducting qubit platforms \cite{Kandala2021,Abrams2020,Li2018,Caldwell2018,MITCoupler}: A flip-flop (also known as XY) Hamiltonian $H_0^{\text{XY}}$ and an XXZ Hamiltonian $H_0^{\text{XXZ}}$ that contains both flip-flop interaction and ZZ (Ising-type) interaction.
\begin{align}\label{XYandXXZ}
    H_0^{\text{XY}}  &= g(\sigma_1^x\sigma_2^x + \sigma_1^y\sigma_2^y) \nonumber \\
    H_0^{\text{XXZ}} &= g(\sigma_1^x\sigma_2^x + \sigma_1^y\sigma_2^y + \eta \sigma_1^z\sigma_2^z )
\end{align}

As an example, we set the parameter $\eta=1/2$ in the following, and similar results are expected for different $\eta$ values. We now perform our gate speed optimization described in Section IV with our experimental static Hamiltonian $H_0$ in Eq.\,\eqref{H0} replaced by the above $H_0^{\text{XY}}$ or $H_0^{\text{XXZ}}$, and the target gate being either SWAP or CNOT.

\begin{figure}
    \includegraphics[width=0.9\columnwidth]{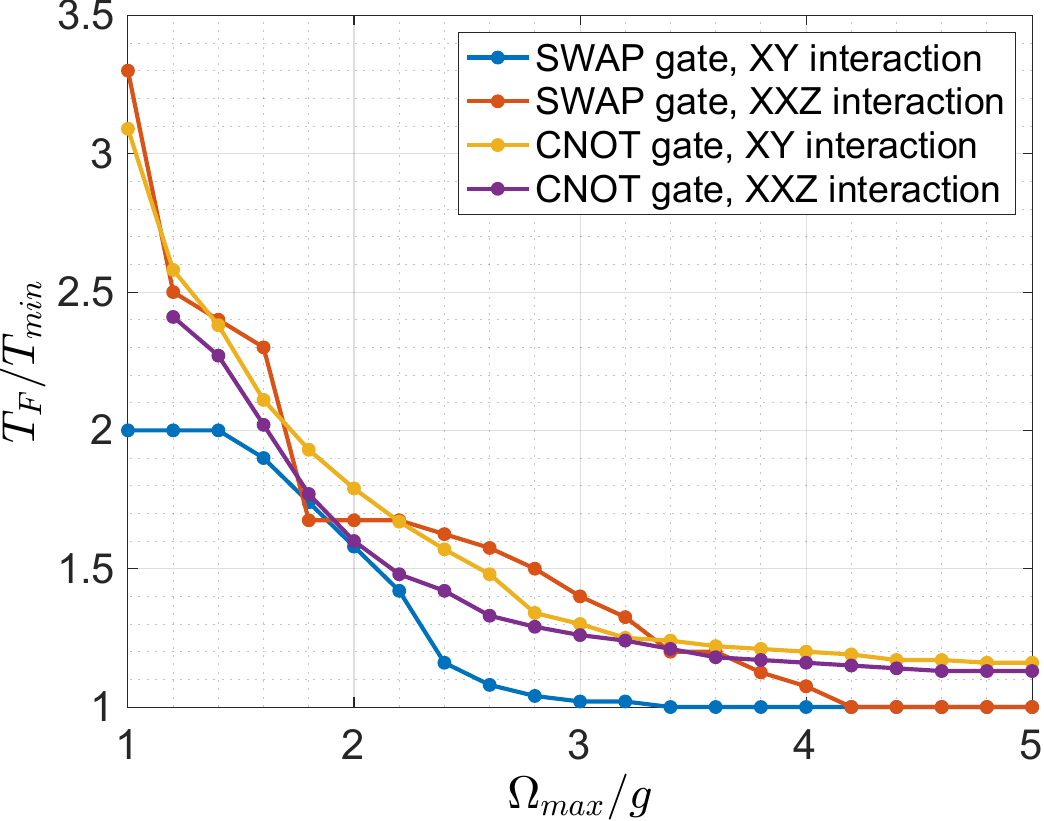}
    \caption{Minimum gate times at different maximum drive strengths achieved by our optimization method for an XY or XXZ interacting Hamiltonian shown in Eq.\,\eqref{XYandXXZ}, with the target gate being SWAP or CNOT. Every point here has average gate fidelity $F>99.99\%$.}
    \label{fig:XYandXXZ}
\end{figure}

In Fig.\,\ref{fig:XYandXXZ}, we show the minimum gate time $T$ (in unit of the corresponding analytical speed limit $T_{\text{min}}$) as a function of the maximum drive strength $\Omega_{\text{max}}$ (in unit of the interaction strength $g$), similar to Fig.\,\ref{fig:OptCNOT}. Note that for the CNOT gate (as well as the CZ gate), $T_{\text{min}}=\pi/(4g)$ holds for $H_0$, $H_0^{\text{XY}}$ and $H_0^{\text{XXZ}}$. But for the SWAP gate, $T_{\text{min}}=3\pi/(8g)$ for $H_0^{\text{XY}}$ and $T_{\text{min}}=3\pi/(10g)$ for $H_0^{\text{XXZ}}$. In other words, the XY or XXZ interaction can generate a SWAP gate faster than the Ising interaction used in our experiment at the same strength.

We have also set a higher fidelity threshold of $F>99.99\%$ to better show the potentials of our optimization method in Fig.\,\ref{fig:XYandXXZ}, while the number of pulse segments, random seeds and iterations remain identical to those in Fig.\,\ref{fig:OptCNOT} (for our experiment such a high fidelity threshold is unnecessary due to experimental error sources). We find that for both $H_0^{\text{XY}}$ and $H_0^{\text{XXZ}}$, one can reach $T\approx T_{\text{min}}$ with $\Omega_{\text{max}}/g<4$. For the CNOT gate, one needs larger drive strengths to reach the theoretical speed limit, but at moderate drive strengths ($\Omega_{\text{max}}\approx 3g$ as in our experiment), one can still get close to the theoretical speed limit ($T\approx 1.2 T_{\text{min}}$).

\vspace{-5pt} 
\section{Process Tomography and SPAM error correction}
\vspace{-5pt} 

\begin{figure*}
    \centering
    \includegraphics[width=0.33\textwidth]{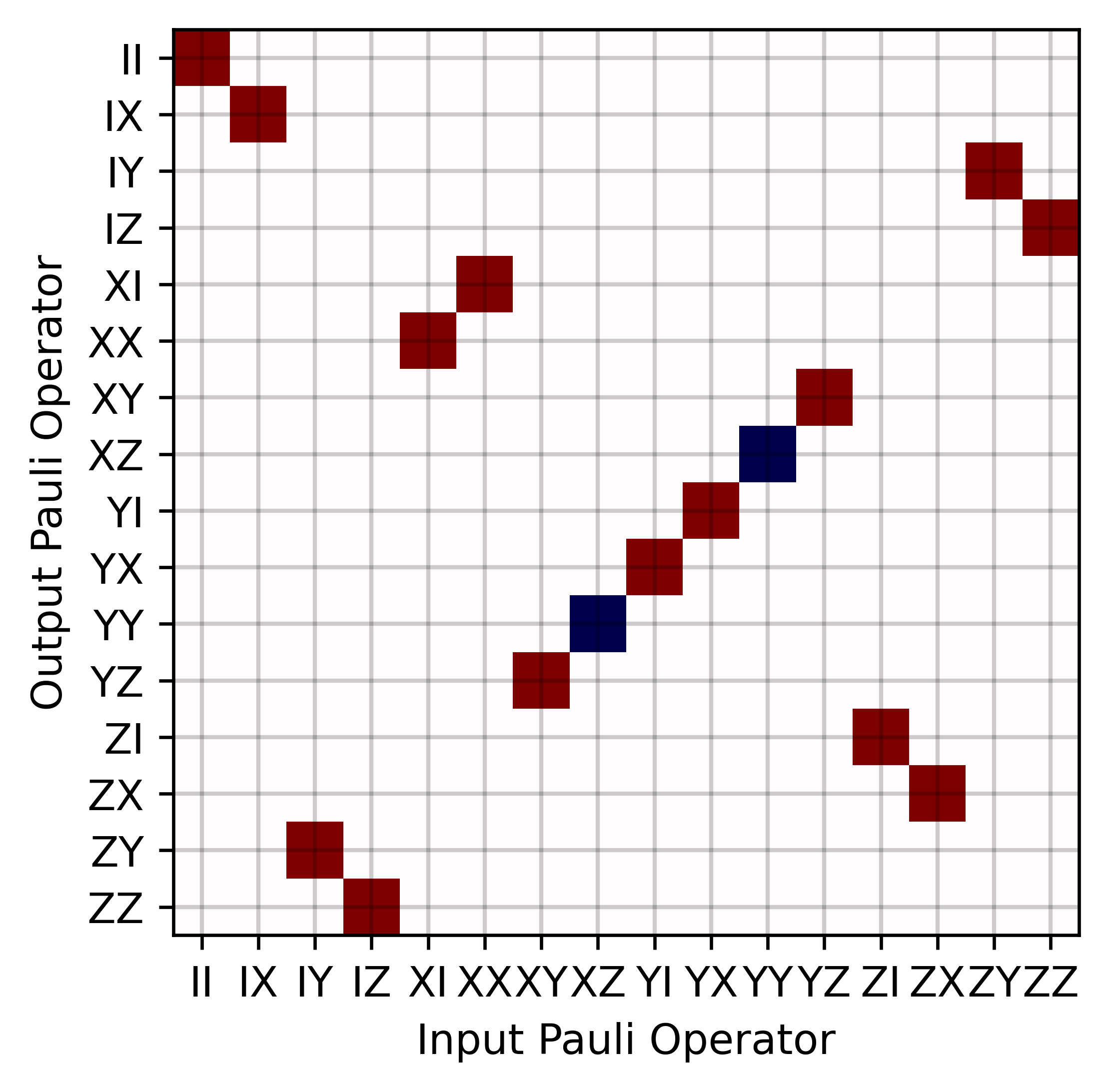}
    \includegraphics[width=0.33\textwidth]{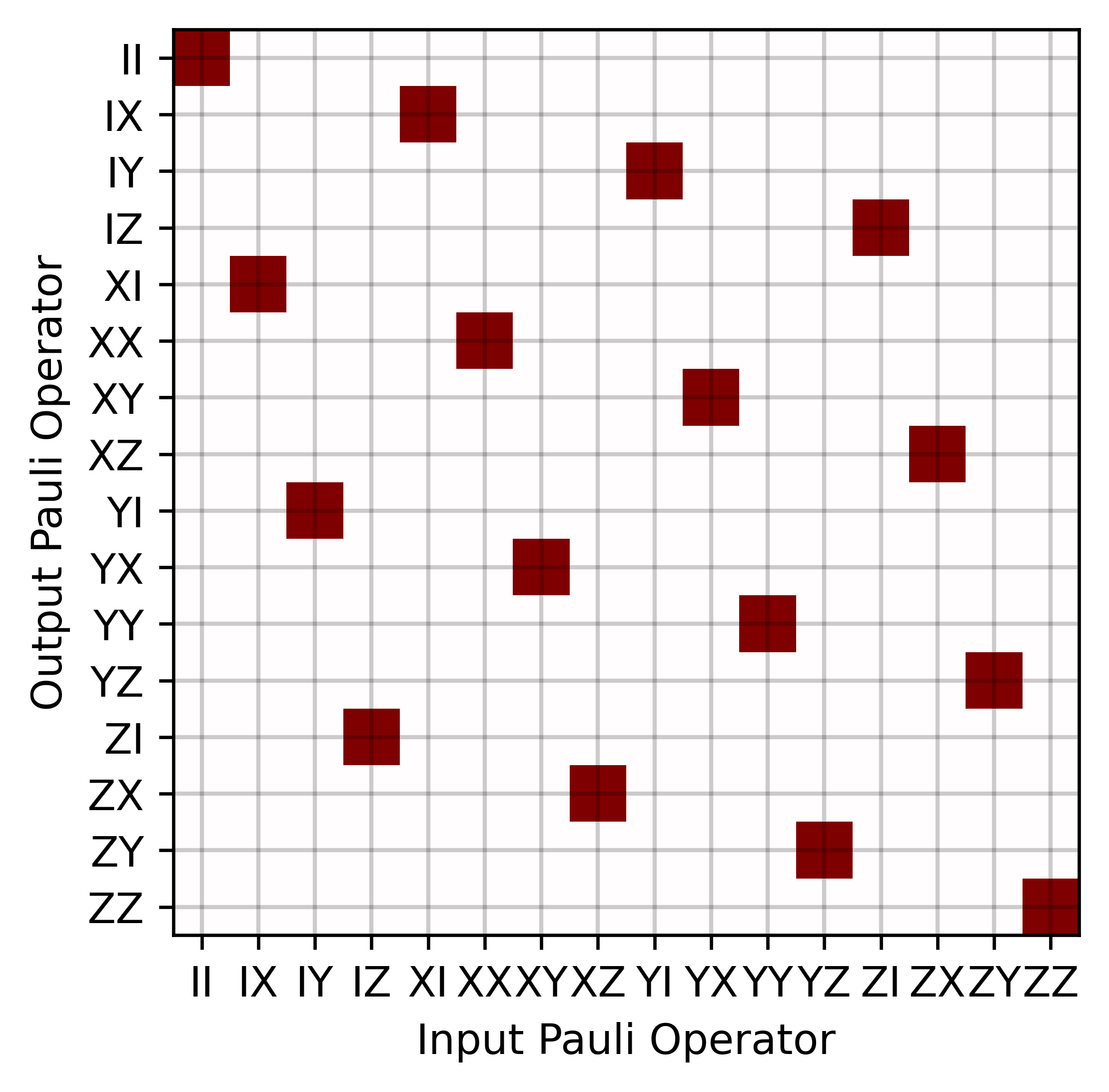}
    \includegraphics[width=0.33\textwidth]{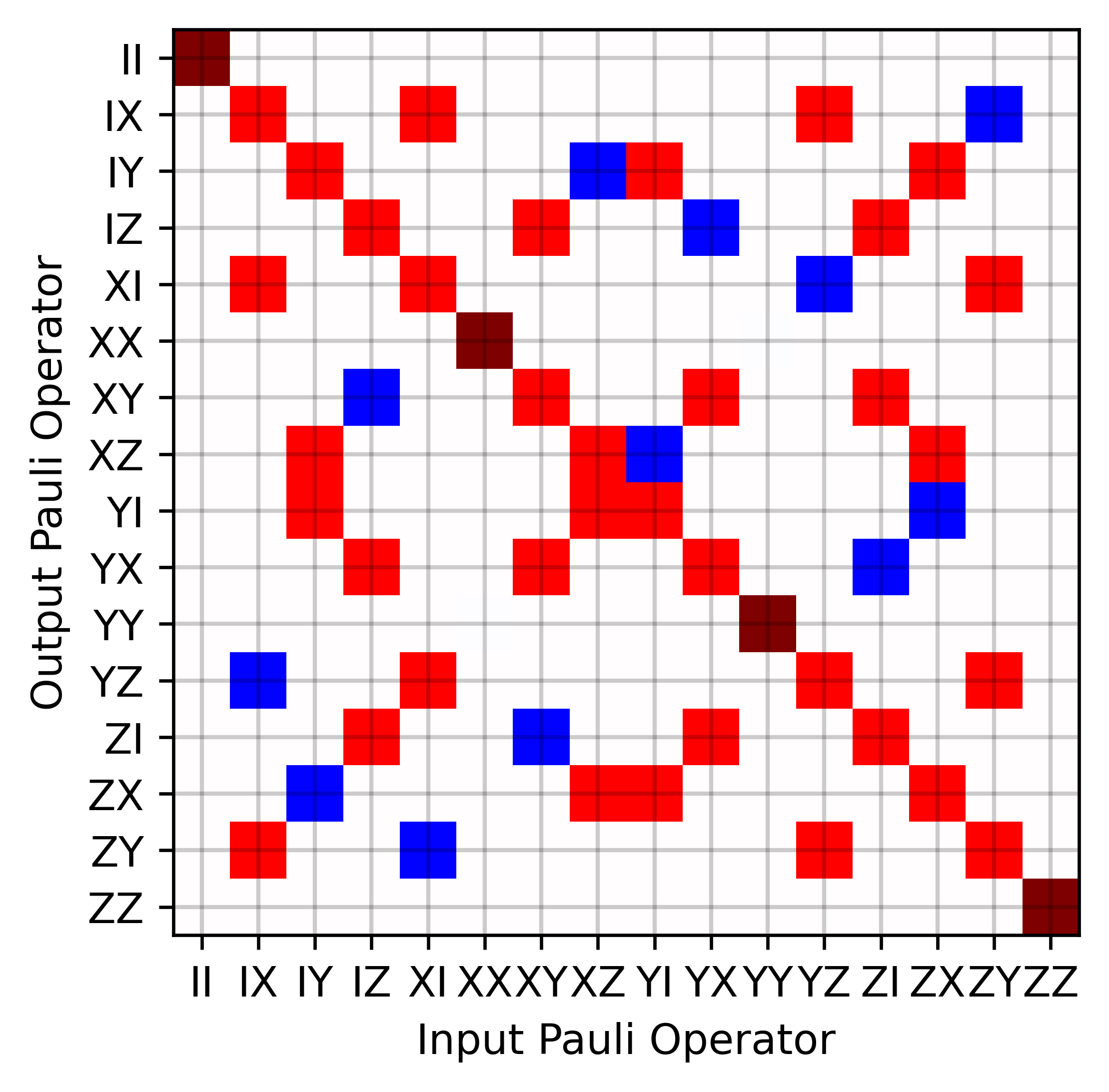}\\
    \includegraphics[width=0.33\textwidth]{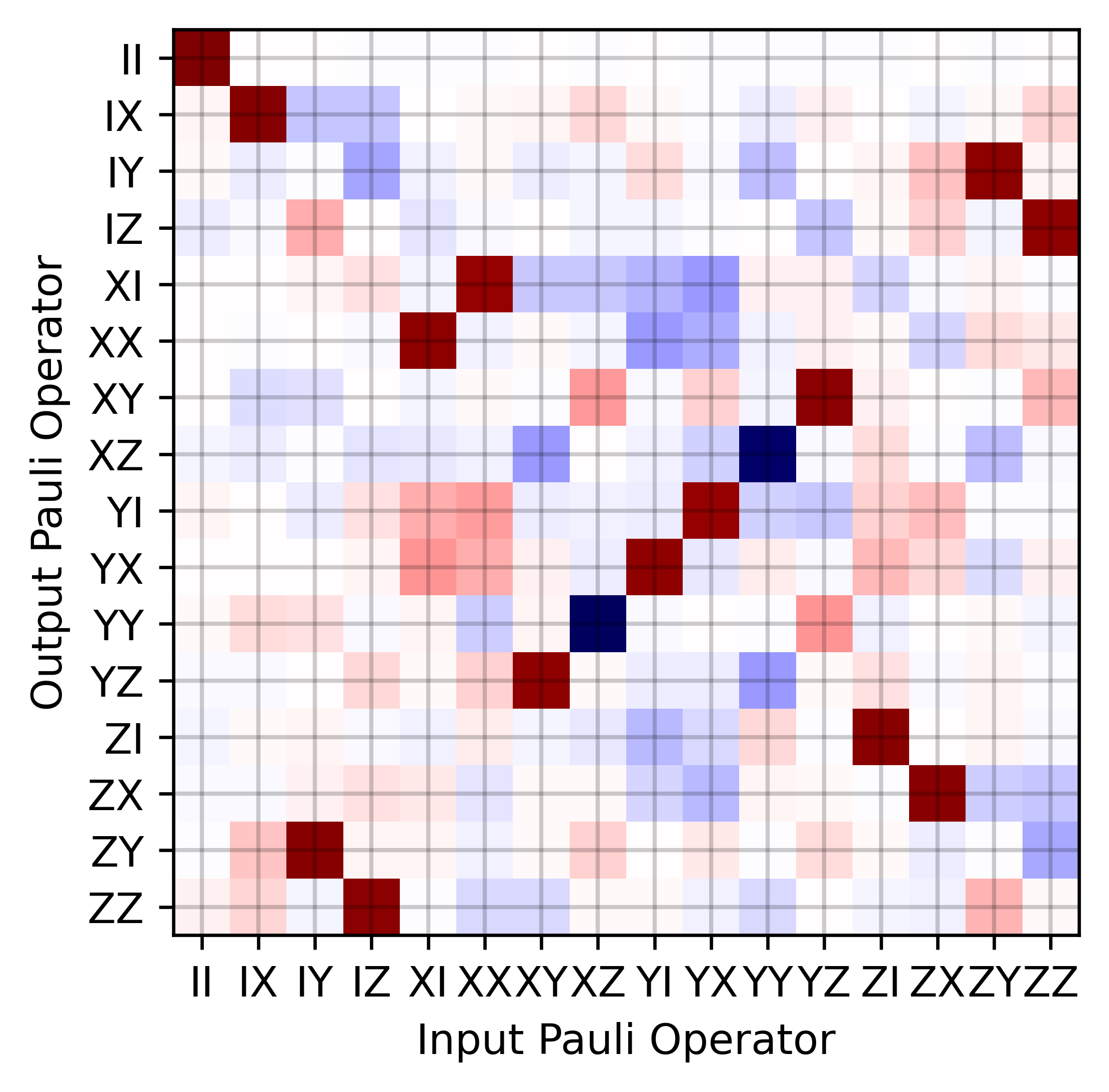}
    \includegraphics[width=0.33\textwidth]{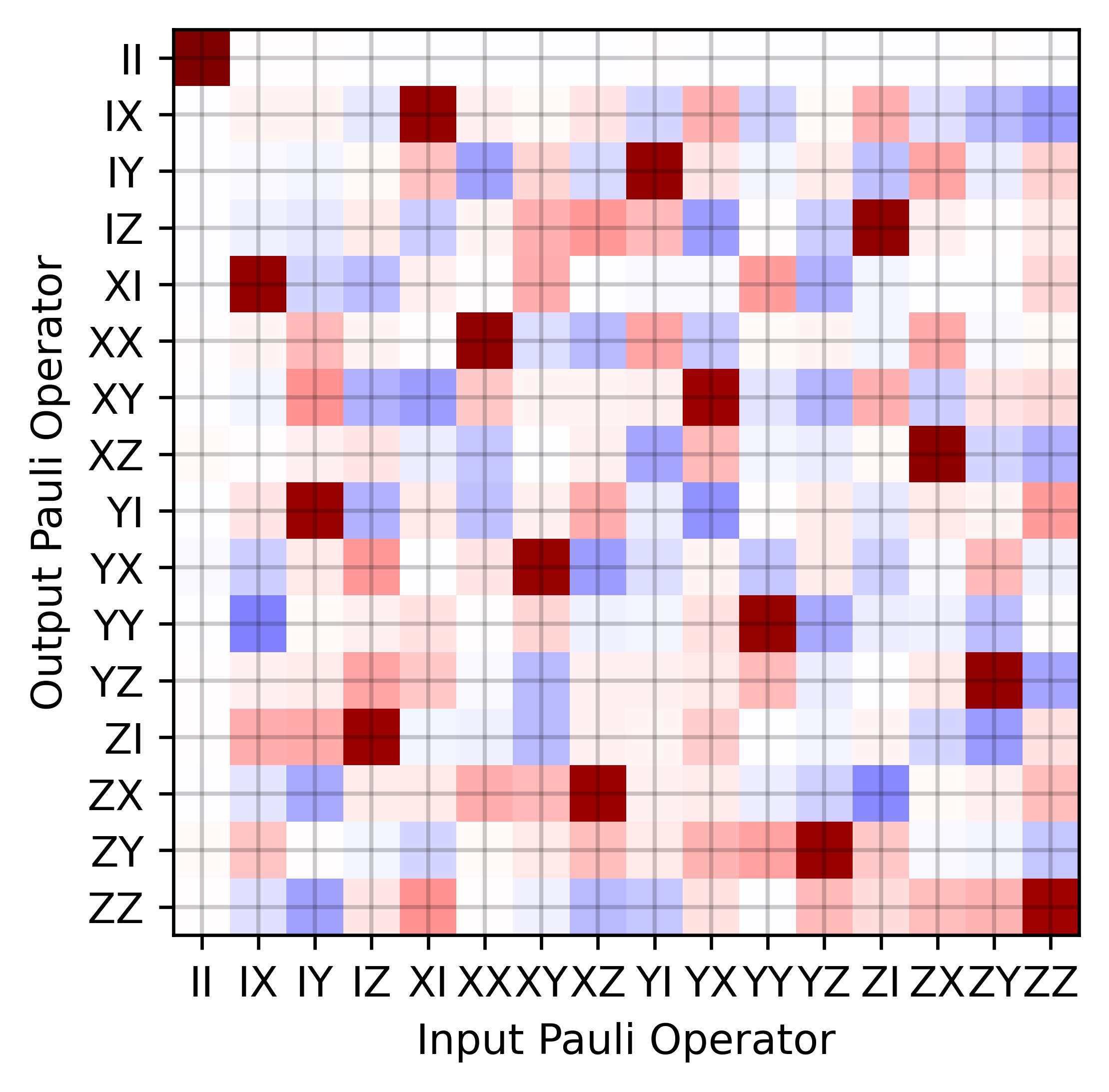}
    \includegraphics[width=0.33\textwidth]{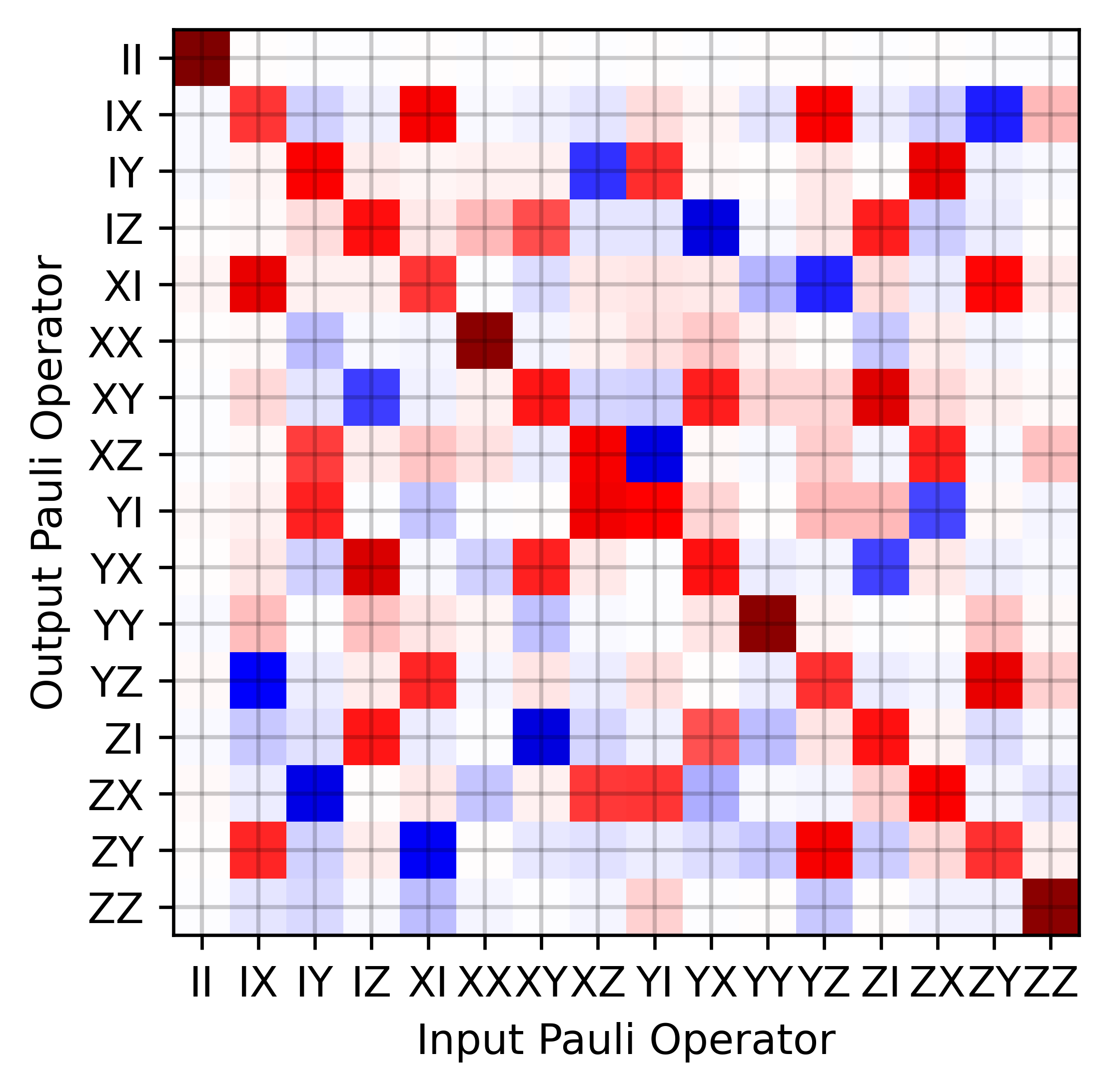}
    \caption{Quantum Process Tomography based on the Pauli transfer matrix for the three target gates we performed experimentally. Top row from left to right: Pauli transfer matrices for an ideal CNOT gate, SWAP gate, and $\sqrt{\text{SWAP}}$ gate. Bottom row from left to right: Examples of SPAM-error-corrected Pauli transfer matrices obtained experimentally for the CNOT , SWAP, and $\sqrt{\text{SWAP}}$ gates, with average gate fidelities being $95.6\%$, $93.1\%$, and $95.7\%$ respectively.}
    \label{fig:QPT}
\end{figure*}

We perform a two-qubit quantum process tomography (QPT) via a standard protocol \cite{Chow2012,GroveTomography} based on the measurement of a Pauli transfer matrix $\mathbf{\mathcal{R}}$ defined as
\begin{align}
    \mathcal{R}_{i,j} \equiv \mbox{Tr}\left[P_i \mathcal{E}(P_j) \right], \quad P_i \in \{I,X,Y,Z\} \otimes \{I,X,Y,Z\}
    \label{PTM}
\end{align}
where $\mathcal{E}(\cdot)$ denotes a quantum map of the process being measured.

For an ideal two-qubit gate $U$, we can calculate the $16\times16$ Pauli transfer matrix $\mathbf{\mathcal{R}}$ numerically using Eq.\,\eqref{PTM} (see Fig.\,\ref{fig:QPT} for examples). To measure $\mathbf{\mathcal{R}}$ experimentally, we need to perform single-qubit rotations before and after the quantum process \cite{QPT}. Here we apply 9 post-rotations $R_k \in \{I, X_{-\pi/2}, Y_{\pi/2}\}^{\otimes 2}$ and 36 pre-rotations $R_l \in \{Y_{\pi/2}, Y_{-\pi/2},X_{-\pi/2}, X_{\pi/2}, I, X_{\pi}\}^{\otimes 2}$ \cite{Chow2012}, for a total of 324 different sequences, each of which is repeated for 500 times to suppress statistical noise in the measurement. The single-qubit rotations here are achieved via the `two-axis gate' protocol \cite{Long:2020}, and they are fined tuned to ensure single-qubit gate fidelities up to $99.1\%$ \cite{Long2021}.

Each single experiment returns a measurement outcome as one of the four two-qubit basis states $\ket{j} \in \{\ket{00},\ket{01},\ket{10},\ket{11}\}$. We then group the outcomes of all 162,000 experiments using a tensor $n_{jkl}$ which counts the number of measurement outcome state $\ket{j}$ for the $k^{th}$ post-rotation and $l^{th}$ pre-rotation. To take into account possible measurement errors, we separately measure a $4\times 4$ confusion matrix $\mathcal{P}$ where $\mathcal{P}_{i,j}$ denotes the probability of obtaining measurement outcome as the state $\ket{i}$ if we initialize the two qubits in state $\ket{j}$.

To infer the Pauli transfer matrix $\mathcal{R}$ from the measurement outcomes tensor $n_{jkl}$, we further invoke a maximum likelihood estimation (MLE) method with the following log-likelihood function of $\mathcal{R}$ \cite{GroveTomography}:
\begin{align}
    \log L(\mathcal{R}) & = \sum_{j,k,l}n_{jkl}\text{log}\left(\sum_{m,n=0}^{15} B_{jklmn}\mathcal{R}_{mn}\right)\\
    B_{jklmn}& = \sum_{m^{\prime},n^{\prime}=0}^{15} \sum_{j^{\prime}=0}^3 \nonumber \\ & \mathcal{P}_{jj^{\prime}} \bra{j^{\prime}} P_{m^{\prime}} \ket{j^{\prime}} \mathcal{R}_k)_{m^{\prime}m}(\mathcal{R}_l)_{n n^{\prime}} \Tr(P_{n^{\prime}} \rho_0)
\end{align}
where $\mathcal{R}_k$ and $\mathcal{R}_l$ are the Pauli transfer matrices for the above-defined post-rotation unitary $R_k$ and pre-rotation unitary $R_l$ respectively. $\rho_0=|00\rangle \langle 00|$ is our initial state and $P_{m^{\prime}},P_{n^{\prime}}$ are defined in Eq.\,\eqref{PTM}. The experimental Pauli transfer matrix $\mathcal{R}$ is then obtained by maximizing $\log L(\mathcal{R})$ under the constraint that $\mathcal{R}$ represents a trace-preserving and completely positive quantum map \cite{Chow2012,GroveTomography}.

Before we perform QPT for the speed-optimized two-qubit gates, we first perform the above QPT procedure for a zero-time evolution to obtain a Pauli transfer matrix $\mathcal{R}_I^{\text{exp}}$, which without state preparation and measurement (SPAM) errors should represent an identity quantum map. The measured $\mathcal{R}_I^{\text{exp}}$ can be used to correct the SPAM errors for a quantum process we perform with nonzero time evolution, whose measured Pauli transfer matrix is denoted by $\mathcal{R}_U^{\text{exp}}$ for a target unitary $U$.

To achieve SPAM error correction, we first obtain the process matrix $\chi_I^{\text{exp}}$ and $\chi_U^{\text{exp}}$ for the corresponding Pauli transfer matrix $\mathcal{R}_I^{\text{exp}}$ and $\mathcal{R}_U^{\text{exp}}$ respectively (by inverting Eq.\,\eqref{chi2R} below) \cite{QPT}. The process matrix for a quantum map $\mathcal{E}$ is defined via $\mathcal{E}(\rho)=\sum_{m,n} \chi_{mn} P_m \rho P_n$ with $P_m,P_n$ defined in Eq.\,\eqref{PTM}. Second, we obtain the SPAM error corrected process matrix $\chi_U^{\text{corrected}}$ for the target process using
\begin{align}
    \chi_U^{\text{corrected}}= T^{-1} (T \chi_U^{\text{exp}} T^{\dagger} -V \chi_I^{\text{exp}} V^{\dagger} + \chi_I^{\text{exp}}) (T^{\dagger})^{-1},
\end{align}
where $T_{mn}=\Tr(P_m P_n U^{\dagger})/4$ and $V_{mn}=\Tr(P_m P_n)/4$ \cite{Han2020}. Finally, we  convert the error corrected process matrix $\chi_U^{\text{corrected}}$ to the error corrected Pauli transfer matrix $\mathcal{R}_U^{\text{corrected}}$ via
\begin{align}
    \mathcal{R}_{ij} = \sum_{m,n=0}^{15}\chi_{mn} \Tr{P_i P_m P_j P_n}. \label{chi2R}
\end{align}
Examples of such error-corrected Pauli transfer matrix $\mathcal{R}_U^{\text{corrected}}$ for $U$ being the CNOT, SWAP, or $\sqrt{\text{SWAP}}$ gate are shown in Fig.\,\ref{fig:QPT}. Finally, $\mathcal{R}_U^{\text{corrected}}$ allows us to find the average gate fidelity $F$ to the target gate $U$ via \cite{NielsenFidelity}
\begin{align}
    F=\frac{4\Tr(\mathcal{R}_U^{\text{corrected}}\mathcal{R}_U^{\text{ideal}})+1}{5}.\label{eq:fidelity}
\end{align}
where $\mathcal{R}_U^{\text{ideal}}$ represents the Pauli transfer matrix of the ideal target gate $U$.

\vspace{-5pt} 
\section{Additional Error Analysis}
\vspace{-5pt}

\begin{figure*}
    \centering
    \includegraphics[width=0.33\textwidth]{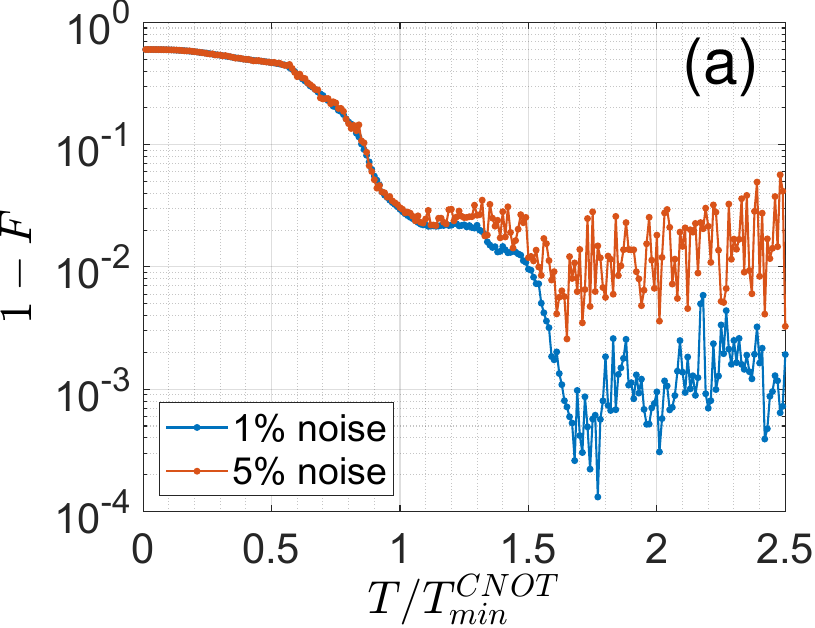}
    \includegraphics[width=0.33\textwidth]{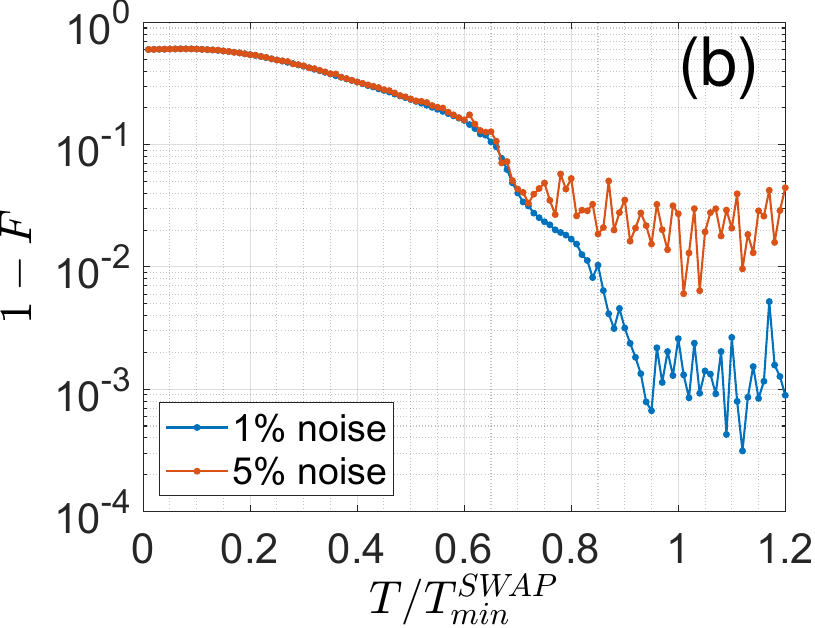}
    \includegraphics[width=0.33\textwidth]{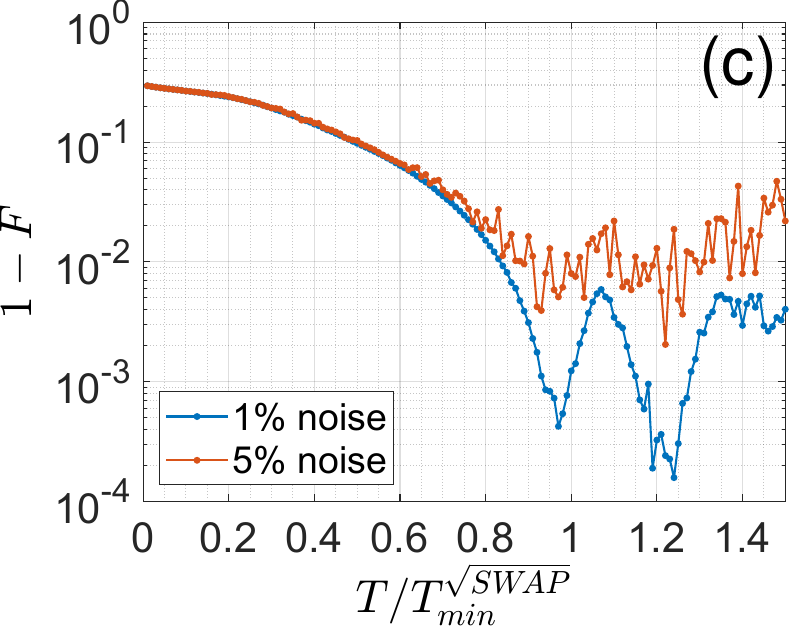}

    \caption{Average gate infidelity $1-F$ calculated using the optimized pulse shapes in Fig.\,3 of the main text but with random Gaussian noise added to each pulse shape parameter $\Omega_{i,m}^{\gamma}$ (see main text) with the target gate being CNOT (a), SWAP (b), and $\sqrt{\text{SWAP}}$ (c). The blue (red) curves correspond to a standard deviation of the Gaussian noise at $0.01\Omega_{\text{max}}$ ($0.05\Omega_{\text{max}}$).}
    \label{fig:noise}
\end{figure*}

One error source shared among all our data points is the statistical error due to quantum measurements. To quantify this error, we simulate additional measurements by adding a Gaussian distributed random noise with zero mean and unity standard deviation on each Pauli operator measured during our QPT \cite{Chow2012}. This allows us to set an upper bound on the statistical error of the mean that would be obtained on re-performing the full experiment with all other error sources held fixed. As shown in Fig.\,3 of the main text, this statistical error on the measured $F$ is less than $1\%$ in all cases.

We have also numerically simulated the effects of imperfect calibration or noises on the optimized pulse shapes (either for amplitudes or phases) by adding random perturbations to each optimized pulse parameter $\Omega_{i,m}^{\gamma}$ (see main text). We expect such perturbations to be present in our experimental setup with magnitudes of a few percent of $\Omega_{\text{max}}$. The simulated average gate fidelity $F$ is shown in Fig.\,\ref{fig:noise}, where all other parameters are identical to the exact $F$ curves in Fig.\,3 of the main text. We see that our optimization method is robust to small amount (1\%) of noises on the pulse shapes, where the fidelity can still exceed $99.9\%$ for gate time close to $T_{\text{min}}$. For larger noises (5\%), the infidelity caused by errors on the drive pulses can be around $1\!-\!2\%$, but such large noises are rare in most experimental platforms.

\bibliographystyle{apsrev4-1}
\bibliography{refs}

\end{document}